\documentclass[12pt,twoside]{article}
\usepackage{amsmath}
\usepackage{amssymb}
\usepackage{array}
\usepackage{graphicx}
\usepackage{geometry}
\usepackage{graphics}
\usepackage{enumerate}
\usepackage[utf8]{inputenc}
\begin{document}
\setcounter{page}{1}
\font\tinyfont=cmr8 \font\headd=cmr8
\pagestyle{myheadings}
\setlength{\columnsep}{1in}
\begin{center}{\bf Modified Models for Neutrino Masses and Mixings}\\
\vskip1cm

Arak M. Mathai\\
Department of Mathematics and Statistics\\
McGill University, Montreal, Canada\\
directorcms458@gmail.com\\

and\\

Hans J. Haubold\\
Office for Outer Space Affairs, United Nations,\\
Vienna International Centre, Vienna, Austria\\
hans.haubold@gmail.com, corresponding author
\end{center}

\vskip.3cm\noindent{\bf Abstract:} The neutrino sector of the seesaw-modified Standard Model is investigated under the anarchy principle. The anarchy principle leading to the seesaw ensemble is studied analytically with tools of random matrix theory. The probability density function is obtained.

\vskip.3cm\noindent{\bf Keywords:} Modified Standard Model, neutrino physics, Dirac neutrino mass matrix, Majorana neutrino mass matrix, probability density functions.

\vskip.3cm\noindent{\bf 1.\hskip.3cm Introduction}

\vskip.5cm\noindent{The Standard Model (SM) of Particle Physics is the pinnacle of the understanding of neutrino physics [1,2]. It comes with a plethora of parameters, the masses and the flavour mixings, that are seemingly not fixed by any known fundamental principle. In the SM, the neutrino spectrum is simple: all neutrinos are massless. Neutrino oscillations, where neutrinos seemingly change flavour in flight, cannot be accommodated in the SM due to the mass of the neutrinos. Neutrino oscillations thus imply massive neutrino eigenstates, and the SM must be extended. Moreover, neutrino oscillation experimental data suggest that the neutrino spectrum is not hierarchical, with three massive light neutrinos and a mixing matrix exhibiting near-maximal mixing [1,2].

To make sense of the neutrino sector, it was argued that the light neutrino mass matrix could be generated randomly from a more fundamental Dirac neutrino mass matrix and a more fundamental Majorana neutrino mass matrix with random elements distributed according to a Gaussian ensemble, a principle dubbed the anarchy principle [3]. These more fundamental neutrino mass matrices would come from the extended SM where the seesaw mechanism occurs [4]. It was argued that the probability density function (PDF) for the mixing angles and phases is the appropriate Haar measure of the symmetry group, implying near-maximal mixings [5,9,12]. We have shown that the PDF can be obtained either by analysing data by diffusion entropy analysis (done for the case of solar neutrinos from observations emanating from SuperKamiokande) [6-8] or as proceeding in this paper.Then, the anarchy principle was analysed mostly numerically, reaching conclusions, for example about the preferred normal hierarchy of the neutrino masses.

Although several numerical results have been obtained, few analytical results on the seesaw ensemble, which is derived from the anarchy principle, exist. That it is the case even though random matrix theory is a well-studied subject in mathematics is surprising. It is therefore clear that a thorough analytical investigation of the seesaw ensemble is possible.

It is an open issue to investigate analytically the seesaw ensemble derived from the anarchy principle with the help of the usual tools of random matrix theory. The seesaw ensemble PDF can be obtained from $N\times N$ fundamental Dirac and Majorana neutrino mass matrices with real or complex elements.  The joint PDF for the singular (eigen) values in the complex (real) case can be derived and it can be shown that the group variables decouple straightforwardly as in the usual Gaussian ensembles.

\vskip.3cm The following notations will be used in this paper: Real scalar ($1\times 1$ matrix) variables, whether random or mathematical, will be denoted by lower-case letters such as $x,y,z$. Real vector ($1\times p$ or $p\times 1$, $p>1$, matrix), matrix ($p\times q$) variables will be denoted by capital letters such as $X,Y$. Variables in the complex domain will be indicated with a tilde such as $\tilde{x},\tilde{y},\tilde{X},\tilde{Y}$. Constants will be denoted by $a,b$ etc for scalars and $A,B$ etc for vectors/matrices.  For a $p\times p$ matrix $B$, $|B|$ or ${\rm det}(B)$ will denote the determinant of the matrix $B$. When $B$ is in the complex domain, one can write $|B|=a+ib,i=\sqrt{(-1)},a,b$ are real scalars. Then, the absolute value of the determinant will be denoted by $|{\rm det}(B)|=|{\rm det}(BB^{*})|^{\frac{1}{2}}=\sqrt{(a^2+b^2)}$ where $B^{*}$ means the conjugate transpose of $B$, that is $B^{*}=(B')^c=(B^c)'$ where a prime denotes the transpose and $c$ in the exponent denotes the conjugate. If $X=(x_{jk})$ is a $p\times q$ matrix in the real domain, where the $x_{jk}$'s are distinct real scalar variables, then the wedge product of their differentials will be denoted as ${\rm d}X=\wedge_{j=1}^p\wedge_{k=1}^q{\rm d}x_{jk}$. If $X=X'$ (real symmetric), then ${\rm d}X=\wedge_{j\ge k}{\rm d}x_{jk}$. For two real scalar variables $x_1$ and $x_2$, the wedge product is defined as ${\rm d}x_1\wedge{\rm d}x_2=-{\rm d}x_2\wedge{\rm d}x_1$ so that ${\rm d}x_j\wedge{\rm d}x_j=0, j=1,2.$ Also, $\int_Xf(X){\rm d}X$ will denote the real-valued scalar function $f(X)$ of $X$ is integrated over $X$. If the $p\times q$ matrix $\tilde{X}$ is in the complex domain, then $\tilde{X}=X_1+iX_2,i=\sqrt{(-1)},X_1,X_2$ are real $p\times q$ matrices, then ${\rm d}\tilde{X}={\rm d}X_1\wedge{\rm d}X_2$. Other notations will be explained whenever they occur for the first time.
\vskip.2cm
This paper is organized as follows: Section 2 gives a mathematical introduction to the models. Section 3 derives the distribution of the light neutrino mass matrix in explicit computable form. Section 4 is providing the densities in terms of the eigenvalues, including the cases of the largest eigenvalue and the smallest eigenvalue.

\vskip.3cm\noindent{\bf 2.\hskip.3cm Modified Dirac and Majorana Neutrino Matrices and Their Distributions}

\vskip.3cm\noindent Let $X$ and $\tilde{X}$ be $p\times n, p\le n$ matrix of rank $p$ in the real and complex domains respectively. If $X$ and $\tilde{X}$ are $p\times n$ matrix-variate random variables in the real and complex domains respectively, then this matrix $X$ corresponds to the $N\times N$ Dirac matrix $M_D$ considered in [5,9,12]. Let $Y>O,\tilde{Y}>O$ be $n\times n$ real positive definite and Hermitian positive definite matrices in the real and complex domains respectively, where $(\cdot)>O$ denotes the matrix $(\cdot)$ is real positive definite or Hermitian positive definite. If $Y$ and $\tilde{Y}$ are $n\times n$ matrix-variate random variables in the real and complex domains, then this $Y$ corresponds to the  $N\times N$ Majorana matrix $M_R$ in [5,9,12]. Let $U=XY^{-1}X'$ where $U$ corresponds to the light neutrino mass matrix $M_{\nu}$ in [5,9,12]. Due to our assumption of $X$ being a full rank matrix, the rows of $X$ are linearly independent so that a singular distribution for any column of $X$ is avoided. The columns of $X$ are $p\times 1$ which corresponds to a $p$-vector in multivariate statistical analysis. Also if the columns of $X$ are iid (independently and identically distributed) then $X$ can represent a sample matrix of a sample of size $n$ from a $p$-variate population. When $p=n$, $X$ will be a $n\times n$ square matrix as considered in [5,9,12]. Let $X$ have a $p\times n$ matrix-variate Gaussian density in the standard form, denoted by $f_1(X)$, where

$$f_1(X){\rm d}X=c_1{\rm e}^{-{\rm tr}(XX')}{\rm d}X,~f_1(\tilde{X}){\rm d}\tilde{X}=\tilde{c}_1{\rm e}^{-{\rm tr}(\tilde{X}\tilde{X}^{*})}{\rm d}\tilde{X}\eqno(1)$$
where $c_1$ is the normalizing constant, $c_1=\pi^{-pn/2}$. If a real scalar scaling constant $b_1>0$ is introduced in the exponent in (1) and write the exponent as $-b_1{\rm tr}(XX')$,
then the normalizing constant changes to $(b_1\pi^{-1} )^{pn/2}$. In the corresponding complex case, $\tilde{c}_1=\pi^{-pn}$ and if a real scaling factor $b_1>0$ is present, then $\tilde{c}_1=(b\pi^{-1} )^{pn}$ respectively. If a location parameter matrix is to be included, then replace $X$ in (1) by $X-M$ and $\tilde{X}$ by $\tilde{X}-\tilde{M}$ where the $M$ and $\tilde{M}$ matrices can act as the mean value or expected value of $X$ and expected value of $\tilde{X}$ respectively, where $M=E[X], \tilde{M}=E[\tilde{X}]$ where $E[(\cdot)]$ denotes the expected value of $(\cdot)$. There is not going to be any change in the normalizing constants $c_1$ and $\tilde{c}_1$. If scaling matrices are also to be inserted, then
replace $XX'$ by $A(X-M)B(X-M)'$ and $\tilde{X}\tilde{X}^{*}$ by $A(\tilde{X}-\tilde{M})B(\tilde{X}-\tilde{M})^{*}$ respectively, where $A>O$ is $p\times p$ and $B>O$ is $n\times n$ real positive definite constant matrices in the real case, and $A=A^{*}>O$ and $B=B^{*}>O$ (Hermitian positive definite matrices) in the complex domain. The normalizing constants will change to $c_1=\pi^{-pn/2}|A|^{\frac{n}{2}}|B|^{\frac{p}{2}}$ and $\tilde{c}_1$ changes to $\tilde{c}_1=\pi^{-pn}|{\rm det}(A)|^{n}|{\rm det}(B)|^{p}$ respectively. These changes are taking place due to Lemma 1 given below. When $XX'$ is changed to $AXBX'$, one can give interpretations in terms of the covariance matrices of the columns and rows of $X$. For example, the inverse of $A$ can act as the covariance matrix of each $p\times 1$ column vector of $X$ and similarly the inverse of $B$ can act as the covariance matrix of each row of $X$ when $X$ is a sample matrix from some $p$-variate population. The above are some of the advantages of considering the $p\times n$ matrix $X$ and inserting location parameter matrix and scaling matrices in $XX'$. Corresponding interpretations can be given in the complex case also.

\vskip.3cm\noindent{\bf Lemma 1.}\hskip.3cm{\it Let $X=(x_{jk})$ be a $p\times q$ matrix in the real domain with $pq$ distinct real scalar variables as elements $x_{jk}$'s. Let $A$ be a $p\times p$ and $B$ be a $q\times q$ real nonsingular constant matrices. Consider the linear transformation $Y=AXB$. Then,
$$Y=AXB,|A|\ne 0,|B|\ne 0\Rightarrow {\rm d}Y=|A|^q|B|^p{\rm d}X$$
In the corresponding complex domain, $\tilde{Y}=A\tilde{X}B$ where $A$ and $B$ are nonsingular and they may be in the real or complex domain, then
$$\tilde{Y}=A\tilde{X}B,|A|\ne 0,|B|\ne 0\Rightarrow {\rm d}\tilde{Y}=|{\rm det}(A)|^{2q}|{\rm det}(B)|^{2p}{\rm d}\tilde{X}.$$}
\vskip.3cm
In the discussion to follow, we will need the Jacobian from a symmetric transformation, for example, $p=q,B=A'$ in the real case in Lemma 1. In the symmetric case, the Jacobian will not be available from Lemma 1.

\vskip.3cm\noindent{\bf Lemma 2.}\hskip.3cm{\it Consider the $p\times p$ matrices $X=X',A$ and $\tilde{X}=\tilde{X}^{*},A$ in the real and complex domains respectively where $A$ is a nonsingular constant matrix and in the complex case $A$ could be  real or complex. Note that we assume $X$ is symmetric in the real domain and $\tilde{X}$ is Hermitian in the complex domain. Then,
$$Y=AXA',|A|\ne 0\Rightarrow {\rm d}Y=|A|^{p+1}{\rm d}X,\tilde{Y}=A\tilde{X}A^{*}\Rightarrow {\rm d}\tilde{Y}=|{\rm det}(A)|^{2p}{\rm d}\tilde{X}.$$}

\vskip.3cm If $X$ is skew symmetric then the exponent $p+1$ changes to $p-1$ in the Jacobian part in Lemma 2. If $\tilde{X}$ is skew Hermitian, there is no change, $2p$ in the exponent will remain as $2p$.
\vskip.2cm Let the $n\times n$ matrix $Y$, corresponding to the Majorana neutrino mass matrix $M_R$ in [5,9,12] have the following density:

$$f_2(Y){\rm d }Y=c_2|Y|^{\alpha-\frac{n+1}{2}}{\rm e}^{-{\rm tr}(Y)}{\rm d}Y,~f_2(\tilde{Y}){\rm d}\tilde{Y}=\tilde{c}_2|{\rm det}(\tilde{Y})|^{\alpha-n}{\rm e}^{-{\rm tr}\tilde{Y}}{\rm d}\tilde{Y}\eqno(2)$$
where $Y=Y'>O$ (real positive definite),  $\tilde{Y}=\tilde{Y}^{*}>O$ (Hermitian positive definite). Here $\alpha$ is a free parameter which may be given some physical interpretations.  The $n\times n$ positive definite matrix $Y$ can always be written as $Y_1Y_1'$ where $Y_1$ is a $n\times n_1,n\le n_1$ matrix of rank $n$ where $n_1$ can be equal to $n$ or $Y_1$ can be a square matrix also. This representation $Y=Y_1Y_1'$ also provides a connection to a Wishart matrix with $\alpha=\frac{n_1}{2}$ in the real case and $\alpha=n$ in the complex  case.  Further, $Y_1$ can be a square root of $Y$ where a square root can be uniquely defined when $Y$ is positive definite or $\tilde{Y}$ is Hermitian positive definite. Thus, the exponents in the densities of $Y$ and $\tilde{Y}$ can have exactly the same structures as in the corresponding densities in [5,9,12]. But in [5,9,12] the densities are Gaussian forms and in (2) above, the densities are real and complex matrix-variate Gamma densities for $Y$ and $\tilde{Y}$. This is the meaning of ``modified'' in the title of the present paper. But, in the complex  case we assume $\tilde{Y}$ to be Hermitian. This is a drawback if someone wishes to take $\tilde{Y}$ in the complex case but restricted to be symmetric only. The normalizing constants $c_2$ and $\tilde{c}_2$ are the following:

$$c_2=\frac{1}{\Gamma_n(\alpha)}, \alpha >\frac{n-1}{2}, \tilde{c}_2=\frac{1}{\tilde{\Gamma}_n(\alpha)}, \alpha >n-1$$
where  $\Gamma_n(\alpha)$ and $\tilde{\Gamma}_n(\alpha)$ are the real and complex matrix-variate gamma functions defined as the following:
\begin{align*}
\Gamma_n(\alpha)&=\begin{cases}\pi^{\frac{n(n-1)}{4}}\Gamma(\alpha)\Gamma(\alpha-\frac{1}{2})...\Gamma(\alpha-\frac{n-1}{2}),\Re(\alpha)>\frac{n-1}{2}\\
\int_{Z>O}|Z|^{\alpha-\frac{n+1}{2}}{\rm e}^{-{\rm tr}(Z)}{\rm d}Z\end{cases}\\
\tilde{\Gamma}_n(\alpha)&=\begin{cases}\pi^{\frac{n(n-1)}{2}}\Gamma(\alpha)\Gamma(\alpha-1)...\Gamma(\alpha-n+1),\Re(\alpha)>n-1\\
\int_{\tilde{Z}>O}|{\rm det}(\tilde{Z})|^{\alpha-n}{\rm e}^{-{\rm tr}(\tilde{Z})}{\rm d}\tilde{Z},\Re(\alpha)>n-1,\end{cases}\tag{3}\end{align*}
where $\Re(\cdot)$ means the real part of $(\cdot)$. Gamma function exists for complex $\alpha$ also under the above conditions. But in statistical problems the parameters are usually real because we wish to restrict a density function to be real-valued scalar function.
Let $Y^{\frac{1}{2}}$ be the positive definite square root of the positive definite matrix $Y>O$. Then, if we consider $X$ of (1) to be scaled by $XY^{-\frac{1}{2}}$ then this scaling has the effect of making the rows of $X$ correlation-free if $Y$ is the correlation matrix of each row of $X$. Thus, in a physical situation if the rows are likely to be correlated then they can be made correlation free by scaling with the proper scaling matrix, namely the square root of the inverse of the correlation matrix. In the scaled $XX'$, namely $AXBX'$ considered above, $B^{-1}$ corresponds to $Y$ in the present discussion. When scaled with the proper scaling matrix $XX'$ goes to $(XY^{-\frac{1}{2}})(XY^{-\frac{1}{2}})'=XY^{-1}X'$ and one has similar changes in the complex case also. Hence, our light neutrino mass matrix $U=XY^{-1}X'$ has proper interpretations in terms of scaling models, making rows correlation free etc.
\vskip.2cm
Our interest is to derive the density of $U$. For this purpose, we need either the assumption that $X$ and $Y$ are independently distributed, in that case the joint density of $X$ and $Y$ is $f_1(X)f_2(Y)$, the product, or we have to assume that $f_1(X)$ is a conditional density, in the sense, for every given $Y$, one has the density of $X$ a matrix-variate Gaussian as in (1) and $f_2(Y)$ is then the marginal density of $Y$ and again the joint density will be the product $f_1(X)f_2(Y)$. We will assume $f_1(X)$ being a conditional density of $X$ for every given $Y$ and derive the density of $U=XY^{-1}X'$ and $\tilde{U}=\tilde{X}\tilde{Y}^{-1}\tilde{X}^{*}$.

\vskip.3cm\noindent{\bf 3.\hskip.3cm Derivation of the Density of $U$}
\vskip.3cm\noindent
In our notation, the light neutrino mass matrix is $U=XY^{-1}X'$ in the real case and $\tilde{U}=\tilde{X}\tilde{Y}^{-1}\tilde{X}^{*}$ in the complex case. Let us consider the real case first.  $U$ can be written as $U=(XY^{-\frac{1}{2}})(XY^{-\frac{1}{2}})'$ since $Y$ is symmetric real positive definite, where, for example, $Y^{\frac{1}{2}}$ is the positive definite square root of the positive definite matrix $Y>O$. Let $Z=XY^{-\frac{1}{2}}\Rightarrow X=ZY^{\frac{1}{2}}$ and ${\rm d}X=|Y|^{\frac{p}{2}}{\rm d}Z$, for fixed $Y$, by using Lemma 1, and $U=ZZ'$. The joint density of $X$ and $Y$, is the conditional density of $X$, given $Y$, times the marginal density of $Y$. That is, denoting the joint density by $f(X,Y)$, we have the following:

\begin{align*}
f(X,Y){\rm d}X\wedge{\rm d}Y&=c_1c_2|Y|^{\alpha-\frac{n+1}{2}}{\rm e}^{-{\rm tr}(ZYZ')}{\rm e}^{-{\rm tr}(Y)}|Y|^{\frac{p}{2}}{\rm d}Z\wedge{\rm d}Y\\
&=c_1c_2|Y|^{\alpha+\frac{p}{2}-\frac{n+1}{2}}{\rm e}^{-{\rm tr}[Y(I_n+Z'Z)]}{\rm d}Z\wedge{\rm d}Y.\end{align*}
But
$${\rm tr}(Y+ZYZ')={\rm tr}(Y(I_n+YZ'Z))={\rm tr}[(I_n+Z'Z)^{\frac{1}{2}}Y(I_n+Z'Z)^{\frac{1}{2}}]$$
Even though $Z'Z$ is singular and positive semi-definite, due to the presence of the identity matrix $I=I_n$, we may take $I_n+Z'Z$ to be positive definite and hence one may consider the positive definite square root of $I_n+Z'Z$. Now, we can integrate out $Y$ by using a real matrix-variate gamma of (3). That is, from Lemma 2
$$\int_{Y>O}|Y|^{\frac{p}{2}+\alpha-\frac{n+1}{2}}{\rm e}^{-{\rm tr}[(I_n+Z'Z)^{\frac{1}{2}}Y(I_n+Z'Z)^{\frac{1}{2}}]}{\rm d}Y=\Gamma_n(\alpha+\frac{p}{2})|I_n+Z'Z|^{-(\alpha+\frac{p}{2})},\alpha >\frac{n-1}{2}.$$
But we can write $|I_n+Z'Z|$ in terms of the $p\times p$ real positive definite matrix $ZZ'$. Consider the expansion of the following determinant in two different ways in terms of its submatrices, denoting the determinant by $\eta$:
\begin{align*}
\eta&=\left\vert\begin{matrix}I_p&-Z\\
Z'&I_n\end{matrix}\right\vert =|I_p|~|I_n-Z'I_p^{-1}(-Z)|=|I_p|~|I_n+Z'Z|=|I_n+Z'Z|\\
\eta&=|I_n|~|I_p-(-Z)I_n^{-1}Z'|=|I_n|~|I_p+ZZ'|=|I_p+ZZ'|\Rightarrow |I_n+Z'Z|=|I_p+ZZ'|.\tag{4}\end{align*}
Hence, the density of $Z$, denoted by $g(Z)$, is the following:
$$g(Z){\rm d}Z=c_1c_2\Gamma_n(\alpha+\frac{p}{2})|I_p+ZZ'|^{-(\alpha+\frac{p}{2})}{\rm d}Z.\eqno(5)$$
Going through steps parallel to the real case, one has the corresponding result in the complex case, denoted by $\tilde{g}(\tilde{Z})$ as the following:

$$\tilde{g}(\tilde{Z}){\rm d}\tilde{Z}=\tilde{c}_1\tilde{c}_2\tilde{\Gamma}_n(\alpha+p)|{\rm det}(I+\tilde{Z}\tilde{Z}^{*})|^{-(\alpha+p)}{\rm d}\tilde{Z}.\eqno(6)$$
Our matrix is $U=ZZ'$. We can go from the density of $Z$ to the density of $U=ZZ'$ by using the following result from [10], see also [2,4], which will be stated as a lemma.

\vskip.3cm\noindent{\bf Lemma 3.}\hskip.3cm{\it Let $X=(x_{jk})$ be a $p\times q,p\le q$ matrix of rank $p$ in the real domain where the $x_{jk}$'s are distinct real scalar variables. Let $S=XX'$. Then, going through a transformation involving a lower triangular matrix with positive diagonal elements and a unique semi-orthonormal matrix and then integrating out the differential element corresponding to the semi-orthonormal matrix, we have the following relationship between ${\rm d}X$ and ${\rm d}S $:
$${\rm d}X=\frac{\pi^{\frac{pq}{2}}}{\Gamma_p(\frac{q}{2})}|S|^{\frac{q}{2}-\frac{p+1}{2}}{\rm d}S.$$
In the corresponding complex case, let $\tilde{X}$ be $p\times q,p\le q$ matrix in the complex domain with distinct scalar complex variables as elements. Let $\tilde{S}=\tilde{X}\tilde{X}^{*}$. Then, going through a transformation involving a lower triangular matrix with real and positive diagonal elements and a unique semi-unitary matrix and then integrating out the differential element corresponding to the semi-unitary matrix, we have the following connection:
$${\rm d}\tilde{X}=\frac{\pi^{pq}}{\tilde{\Gamma}_p(q)}|{\rm det}(\tilde{S})|^{q-p}{\rm d}\tilde{S}.$$}

\vskip.3cm With the help of Lemma 3, we can go to the density of $Z$ in (5) to the density of $U=ZZ'$, denoted by $g_1(U)$. Since the variable is changed from a $p\times n$ matrix to a $p\times p$ matrix, the normalizing constant will change. Hence we may write

$$g_1(U){\rm d}U=c|U|^{\frac{n}{2}-\frac{p+1}{2}}|I+U|^{-(\alpha+\frac{p}{2})}{\rm d}U\eqno(7)$$
for $\alpha >\frac{n-1}{2},n>p-1$, where $c$ is the corresponding normalizing constant. This $g_1(U)$ is a real matrix-variate type 2 beta density with the parameters $(\alpha +\frac{p-n}{2},\frac{n}{2})$. Hence, the normalizing constants, denoted by $c$ in the real case and $\tilde{c}$ in the complex case, are the following:

$$c^{-1}=\frac{\Gamma_p(\frac{n}{2})\Gamma_p(\alpha+\frac{p}{2}-\frac{n}{2})}{\Gamma_p(\alpha+\frac{p}{2})},\alpha >\frac{n-1}{2},
\tilde{c}^{-1}=\frac{\tilde{\Gamma}_p(n)\tilde{\Gamma}_p(\alpha+p-n)}{\tilde{\Gamma}_p(\alpha+p)},\alpha >n-1,\eqno(8)$$
evaluated from  real and complex $p\times p$ matrix-variate type 2 beta densities respectively. Using steps parallel to the real case, we have the corresponding density $\tilde{g}_1(\tilde{U})$ in the complex case as the following, for $\alpha >n-1$:

$$\tilde{g}_1(\tilde{U}){\rm d}\tilde{U}=\tilde{c}|{\rm det}(\tilde{U})|^{n-p}|{\rm de}(I+\tilde{U})|^{-(\alpha+p)}{\rm d}\tilde{U}\eqno(9)$$
where $\tilde{c}$ is given in (8) and $\tilde{U}$ has a type-2 beta density with the parameters $n$ and $\alpha+p-n$. Observe that if $Z_1$ and $Z_2$ are statistically independently distributed $p\times p$ matrix-variate gamma random variables with the same scale parameter matrix, including Wishart variables, then $U_1=(Z_1+Z_2)^{-\frac{1}{2}}Z_1(Z_1+Z_2)^{-\frac{1}{2}}$ is matrix-variate type 1 beta distributed and $U_2=Z_2^{-\frac{1}{2}}Z_1Z_2^{-\frac{1}{2}}$ is type 2 beta  distributed, including F-distribution. Further, $U=XY^{-1}X'$ and $U_3=X'XY^{-1}$ have the same nonzero eigenvalues, and when $p=n$, $X'X$ is Wishart-distributed since our $X$ is assumed to be standard normal. Thus, our $U$ has the structure giving rise to the same eigenvalues as that of a type 2 beta matrix.

\vskip.3cm\noindent{\bf 4.\hskip.3cm Densities in terms of the Eigenvalues}

\vskip.3cm\noindent
We can convert $U$ and $\tilde{U}$ of (7) and (9) and write the densities in terms of their eigenvalues. If $\mu_j$ is an eigenvalue of $U$, then $0<\mu_j<\infty,j=1,...,p$. Similar is the case for the eigenvalues of $\tilde{U}$. For convenience, let us convert $U$ and $\tilde{U}$ to the corresponding type 1 beta form. Consider the transformation
$$V=(I+U)^{-\frac{1}{2}}U(I+U)^{-\frac{1}{2}},\tilde{V}=(I+\tilde{U})^{-\frac{1}{2}}\tilde{U}(I+\tilde{U})^{-\frac{1}{2}}.\eqno(10)$$
Then, $V$ and $\tilde{V}$ will be $p\times p$ matrix-variate type 1 beta with the same parameters, see [10,11]. Let the densities of $V$ and $\tilde{V}$ be denoted by  $g_2(V)$ and $\tilde{g}_2(\tilde{V})$ respectively. Then,
$$g_2(V){\rm d}V=\frac{\Gamma_p(\alpha+\frac{p}{2})}{\Gamma_p(\frac{n}{2})\Gamma_p(\alpha+\frac{p}{2}-\frac{n}{2})}
|V|^{\frac{n}{2}-\frac{p+1}{2}}|I-V|^{\alpha+\frac{p}{2}-\frac{n}{2}-\frac{p+1}{2}}{\rm d}V\eqno(11)$$
and
$$\tilde{g}_2(\tilde{V}){\rm d}\tilde{V}=\frac{\tilde{\Gamma}_p(\alpha+p)}{\tilde{\Gamma}_p(n)\tilde{\Gamma}_p(\alpha+p-n)}|{\rm det}(\tilde{V})|^{n-p}|{\rm det}(I-\tilde{V})|^{\alpha-n}{\rm d}\tilde{V}\eqno(12)$$for $\alpha >\frac{n-1}{2},n-1$ respectively in the real and the corresponding complex case. If $\lambda_j$ is an eigenvalue of $V$, then $\lambda_j=\frac{\mu_j}{(1+\mu_j)}\Rightarrow \mu_j=\frac{\lambda_j}{(1-\lambda_j)},0<\lambda_j<1,0<\mu_j<\infty,j=1,...,p.$ Let $Q$ be a $p\times p$ unique orthonormal matrix, $QQ'=I,Q'Q=I$ such that $Q'VQ={\rm diag}(\lambda_1,...,\lambda_p)$ with $1>\lambda_1>\lambda_2>...>\lambda_p>0.$ Uniqueness for $Q$ can be achieved, for example, by restricting the elements in the leading diagonal positions to be positive, multiply rows by $-1$ when necessary. Correspondingly, let $\tilde{Q}$ be a unique unitary matrix, $\tilde{Q}\tilde{Q}^{*}=I,\tilde{Q}^{*}\tilde{Q}=I$ such that $\tilde{Q}^{*}\tilde{V}\tilde{Q}={\rm diag}(\lambda_1,...,\lambda_p)$, where $\tilde{Q}^{*}$ means the conjugate transpose of $\tilde{Q}$. When $\lambda_j$'s are real scalar variables we can assume $Pr\{\lambda_i=\lambda_j,i\ne j\}=0$ almost surely. Hence, without loss of generality we assume that the $\lambda_j$'s are distinct, $1>\lambda_1>...>\lambda_p>0$. Observe that the eigenvalues of Hermitian matrices are also real and hence the eigenvalues of both $V$ and $\tilde{V}$ will be real and we will denote them by the same symbols $\lambda_j$'s. Also, $Q'VQ=D={\rm diag}(\lambda_1,...,\lambda_p)\Rightarrow V=QDQ',|V|=\lambda_1...\lambda_p, |I-V|=\prod_{j=1}^p(1-\lambda_j)$ and when $V$ is transformed to its eigenvalues  factors $\prod_{i<j}(\lambda_i-\lambda_j)$ and $\prod_{i<j}(\lambda_i-\lambda_j)^2$ come in, in the real and complex cases respectively, see [10,11]. If the differential elements corresponding to $Q$ and $\tilde{Q}$ are denoted by ${\rm d}G$ and ${\rm d}\tilde{G}$ respectively, then from [10,11], $G=Q'({\rm d}Q),\tilde{G}=\tilde{Q}^{*}({\rm d}\tilde{Q})$ where, for example, $({\rm d}Q)$ is the matrix of differentials in $Q$ and  the integrals over ${\rm d}G$ and ${\rm d}\tilde{G}$ are the following results which will be written as a lemma, see [10,11]:

\vskip.3cm\noindent{\bf Lemma 4.}\hskip.3cm{\it For the $G, {\rm d}G,\tilde{G},{\rm d}\tilde{G}$ as defined above, we have
$$\int{\rm d}G=\frac{\pi^{\frac{p^2}{2}}}{\Gamma_p(\frac{p+1}{2})}, \int{\rm d}\tilde{G}=\frac{\pi^{p(p-1)}}{\tilde{\Gamma}_p(p)}.$$}

\vskip.3cm Let us verify this lemma for $p=2,3$. For a $p\times p$ real positive definite matrix $X$ we have
$$\int_{X>O}|X|^{\alpha-\frac{p+1}{2}}{\rm e}^{-{\rm tr}(X)}{\rm d}X=\Gamma_p(\alpha),\alpha >\frac{p-1}{2}$$
from the real matrix-variate gamma integral. In the complex case, let the $p\times p$ matrix $\tilde{X}$ be Hermitian positive definite. Then, from the complex matrix-variate gamma integral we have
$$\int_{\tilde{X}>O}|{\rm det}(\tilde{X})|^{\alpha-p}{\rm e}^{-{\rm tr}(\tilde{X})}{\rm d}\tilde{X}=\tilde{\Gamma}_p(\alpha),\alpha >p-1.$$
Consider the integrals in the real and complex cases when $\alpha=\frac{p+1}{2}$ in the real case and $\alpha =p$ in the complex case. Then,
$$\int_{X>O}{\rm e}^{-{\rm tr}(X)}{\rm d}X=\Gamma_p(\frac{p+1}{2}),\int_{\tilde{X}>O}{\rm e}^{-{\rm tr}(\tilde{X})}{\rm d}\tilde{X}=\tilde{\Gamma}_p(p).$$
If we go through a unique orthonormal transformation involving an orthonormal matrix $Q$ then in the real case
$$\Gamma_p(\frac{p+1}{2})=\int_D\{\prod_{i<j}(\lambda_i-\lambda_j)\}{\rm e}^{-{\rm tr}(D)}{\rm d}D\int_{Q}{\rm d}G$$
and in the corresponding complex case
$$\tilde{\Gamma}_p(p)=\int_D\{\prod_{i<j}(\lambda_i-\lambda_j)^2\}{\rm e}^{-{\rm tr}(D)}{\rm d}D\int_{\tilde{Q}}{\rm d}\tilde{G}.$$
Then, in the real case, for $p=2, \Gamma_p(\frac{p+1}{2})=\Gamma_2(\frac{3}{2})=\pi/2.$ $\int_Q{\rm d}G=\frac{\pi^{\frac{p^2}{2}}}{\Gamma_p(\frac{p}{2})}  =\frac{p^2}{\Gamma_2(1)}=\pi$. Now, $\Gamma_p(\frac{p+1}{2})$ divided by $\int_Q{\rm d}G=\pi$ from Lemma 4 gives $\frac{1}{2}$. Now, consider the integral over $D$ for $p=2$ in the real case. Let $u_1=\lambda_1-\lambda_2$.
\begin{align*}
\int_D(\lambda_1-\lambda_2){\rm e}^{-(\lambda_1+\lambda_2)}{\rm d}D&=\int_{u_1=0}^{\infty}u_1{\rm e}^{-u_1}{\rm d}u_1\int_{\lambda_2=0}^{\infty}{\rm e}^{-2\lambda_2}{\rm d}\lambda_2\\
&=\frac{1}{2}.\end{align*}
Hence, Lemma 4 for $p=2$ in the real case is verified. Now, for $p=3$ in the real case, the left side quantity
$\Gamma_p(\frac{p+1}{2})=\Gamma_3(2)= \frac{\pi^2}{2}$. $\int_Q{\rm d}G=\frac{\pi^\frac{p^2}{2}}{\Gamma_p(\frac{p}{2})}=\frac{\pi^{9/2}}{\Gamma_3(\frac{3}{2})}=2\pi^2$. Then, $\Gamma_p(\frac{p+1}{2})/\int_Q{\rm d}G=\frac{\pi^2}{2}/(2\pi^2)=\frac{1}{4}$. Now, consider the integral over $D$. Let $u_1=\lambda_1-\lambda_2),u_2=\lambda_2-\lambda_3,u_3=\lambda_3$. Then,
\begin{align*}
&\int_D(\lambda_1-\lambda_2)(\lambda_1-\lambda_3)(\lambda_2-\lambda_3){\rm e}^{-(\lambda_1+\lambda_2+\lambda_3)}{\rm d}D\\
&=\int_0^{\infty}\int_0^{\infty}\int_0^{\infty}u_1(u_1+u_2)u_2{\rm e}^{-(u_1+2u_2+3u_3)}{\rm d}u_1\wedge{\rm d}u_2\wedge{\rm d}u_3\\
&=\frac{1}{3}[\frac{2}{4}+\frac{2}{8}]=\frac{1}{4}.\end{align*}
Hence, Lemma 4 for $p=3$ in the real case is verified. Now, consider the complex case.
$\tilde{\Gamma}_p(p)=\tilde{\Gamma}_2(2)=\pi^{\frac{p(p-1)}{2}}\Gamma(2)\Gamma(1)=\pi$ for $p=2$ and $2\pi^3$ for $p=3$.  $\int_{\tilde{Q}}{\rm d}\tilde{G}=\frac{\pi^{p(p-1)}}{\tilde{\Gamma}_p(p)}=\pi$ for $p=2$ and $\frac{\pi^3}{2}$ for $p=3$. Now, $\tilde{\Gamma}_p(p)/\int_{\tilde{Q}}{\rm d}\tilde{G}$ gives the following:
$\pi/\pi =1$ for $p=2$ and $(2\pi^3)/(\frac{\pi^3}{2})=4$ for $p=3$. Now, consider the integral over $D$ in the complex case. As before, let $u_1=\lambda_1-\lambda_2,u_2=\lambda_2-\lambda_3$. Then, for $p=2$,
\begin{align*}
&\int_D(\lambda_1-\lambda_2)^2{\rm e}^{-(u_1+2\lambda_2)}{\rm d}u_1\wedge{\rm d}\lambda_2\\
&=\int_0^{\infty}\int_0^{\infty}u_1^2{\rm e}^{-(u_1+2\lambda_2)}{\rm d}u_1\wedge{\rm d}\lambda_2=1.
\end{align*} Thus, the result for $p=2$ is verified. Now, consider $p=3$.
\begin{align*}
&\int_0^{\infty}\int_0^{\infty}\int_0^{\infty}u_1^2(u_1+u_2)^2u_2^2{\rm e}^{-(u_1+2u_2+3\lambda_3)}{\rm d}u_1\wedge{\rm d}u_2\wedge{\rm d}\lambda_3\\
&=\int_0^{\infty}{\rm e}^{-3\lambda_3}{\rm d}\lambda_3\int_0^{\infty}\int_0^{\infty}
(u_1^4u_2^2+2u_1^3u _2^3+u_1^2u_2^4){\rm e}^{-u_1-2u_2}{\rm d}u_1\wedge{\rm d}u_2\\
&=4.
\end{align*} Hence, Lemma 4 for $p=3$ in the complex case is verified.
\vskip.2cm
The joint density of $\lambda_1,...,\lambda_p$ is the following, denoted by $g_3(D)$ in the real case and $\tilde{g}_3(D)$ in the complex case:
\begin{align*}
g_3(D){\rm d}D&=\frac{\Gamma_p(\alpha+\frac{p}{2})}{\Gamma_p(\frac{n}{2})\Gamma_p(\alpha+\frac{p}{2}-\frac{n}{2})}\frac{\pi^{\frac{p^2}{2}}}{\Gamma_p(\frac{p}{2})}\{\prod_{i<j}(\lambda_i-\lambda_j)\}\\
&\times \{\prod_{j=1}^p\lambda_j^{\frac{n}{2}-\frac{p+1}{2}}\}\{\prod_{j=1}^p(1-\lambda_j)^{\alpha-\frac{n+1}{2}}\}{\rm d}D\tag{13}\\
\tilde{g}_3(D){\rm d}D&=\frac{\tilde{\Gamma}_p(\alpha+p)}{\tilde{\Gamma}_p(n)\tilde{\Gamma}_p(\alpha+p-n)}\frac{\pi^{p(p-1)}}{\tilde{\Gamma}_p(p)}\{\prod_{i<j}(\lambda_i-\lambda_j)^2\}\\
&\times \{\prod_{j=1}^p\lambda_j^{n-p}\}\{\prod_{j=1}^p(1-\lambda_j)^{\alpha-n}\}{\rm d}D.\tag{14}\end{align*}
\vskip.2cm
Distributions of eigenvalues were considered by many authors starting from the 1930's. A brief history is given in [13]. Different procedures for deriving the densities and different types of representations of the densities of the largest and smallest eigenvalues of gamma matrix, including Wishart matrix, type 1 beta matrix, type 2 beta matrix, including F-matrix, are available in the literature. The discussion in the present paper is based on a procedure developed by the first author in 2020 and reported in [11,13] and it is believed that the representations of the densities of the eigenvalues given here are easier to arrive at for all cases of gamma, type 1 beta and type 2 beta matrices, explicit and amenable to computation.
\vskip.3cm

We can write $\prod_{i<j}(\lambda_i-\lambda_j)$ in the real case and $\prod_{i<j}(\lambda_i-\lambda_j)^2$ in the complex case, in the $p\times p$ matrix case, in terms of Vandermonde's determinant.
$$\prod_{i<j}(\lambda_i-\lambda_j)=\left\vert\begin{matrix}\lambda_1^{p-1} &\lambda_1^{p-2}&...&\lambda_1&1\\
\lambda_2^{p-1}&\lambda_2^{p-2}&...&\lambda_2&1\\
\vdots&\vdots&...&\vdots&\vdots\\
\lambda_p^{p-1}&\lambda_p^{p-2}&...&\lambda_p&1\end{matrix}\right\vert =|A|, A=(a_{ij}),a_{ij}=\lambda_i^{p-j}$$
for all $i$ and $j$. Let us use the general expansion for a determinant. Then, for $K=(k_1,...,k_p)$ where $k_1,...,k_p$ is a given permutation of $1,2,...,p$, we have the following:
$$|A|=\sum_K(-1)^{\rho_K}a_{1k_1}...a_{pk_p}=\sum_K(-1)^{\rho_K}\lambda_1^{p-k_1}\lambda_2^{p-k_2} ...\lambda_p^{p-k_p}.$$
Here, $\rho_K$ is the number of transpositions needed to bring $(k_1,...,k_p)$ to the natural order $(1,2,...,p)$. Then, if $\rho_K$ is odd then we have $-1$ and if $\rho_K$ is even then we have $+1$ as the coefficient. For example, for $p=3$ the possible permutations are $(1,2,3),(1,3,2),(2,3,1),(2,1,3),(3,1,2),(3,2,1)$. There are $3!=6$ terms. In the general case there are $p!$ terms. For example, for the sequence $(1,2,3)$ we have $k_1=1,k_2=2,k_3=3\Rightarrow \rho_K=0$ and the corresponding sign is $+1$. For $(1,3,2)$ we have $k_1=1,k_2=3,k_3=2$. Here one transposition is needed to bring to the natural order $(1,2,3)$ and hence $\rho_K=1$ and the corresponding sign is $-1$, and so on. In the complex case,
$$\prod_{i<j}(\lambda_i-\lambda_j)^2=|A|^2=|AA'|=|A'A|.$$
Let $A'=[\beta_1,\beta_2,...,\beta_p]$, where $\beta_j$ is the $j$-th column of $A'$, $\beta_j'=[\lambda_j^{p-1},\lambda_j^{p-2},...,\lambda_j,1], j=1,...,p$. Let $A'A=B=(b_{ij}),b_{ij}=\sum_{r=1}^p\lambda_r^{2p-(i+j)}$. Then,
$$\prod_{i<j}(\lambda_i-\lambda_j)^2=|B|=|A'A|=\sum_K(-1)^{\rho_K}b_{1k_1}b_{2k_2}...b_pk_p$$
where

\begin{align*}
b_{1k_1} &=\lambda_1^{2p-(1+k_1)}+\lambda_2^{2p-(1+k_1)}+...+\lambda_p^{2p-(1+k_1)} \\
b_{2k_2} &=\lambda_1^{2p-(2+k_2)}+\lambda_2^{2p-(2+k_2)}+...+\lambda_p^{2p-(2+k_2)}\\
\vdots &=\vdots\\
b_{pk_p} &=\lambda_1^{2p-(p+k_p)}+\lambda_2^{2p-(p+k_p)}+...+\lambda_p^{2p-(p+k_p)}
\end{align*}
Let
$$ b_{1k_1}b_{2k_2}...b_{pk_p}=\sum_{r_1,...,r_p}\lambda_1^{r_1}...\lambda_p^{r_p}.\eqno(15)$$
In the real case, the joint density of $\lambda_1,...,\lambda_p$ is the following:
\begin{align*}
g_3(D){\rm d}D&=\frac{\Gamma_p(\alpha+\frac{p}{2})}{\Gamma_p(\frac{n}{2})\Gamma_p(\alpha+\frac{p}{2}-\frac{n}{2})}\frac{\pi^{\frac{p^2}{2}}}{\Gamma_p(\frac{p}{2})}\{\prod_{j=1}^p\lambda_j^{\frac{n}{2}-\frac{p+1}{2}}\}\{\prod_{j=1}^p(1-\lambda_j)^{\alpha-\frac{n+1}{2}}\}\\
&\times (\sum_K(-1)^{\rho_K}\lambda_1^{p-k_1}\lambda_2^{p-k_2}...\lambda_p^{p-k_p}){\rm d}D\\
&=\frac{\Gamma_p(\alpha+\frac{p}{2})}{\Gamma_p(\frac{n}{2})\Gamma_p(\alpha+\frac{p}{2}-\frac{n}{2})}\frac{\pi^{\frac{p^2}{2}}}{\Gamma_p(\frac{p}{2})}\sum_K(-1)^{\rho_K}\lambda_1^{m_1}...\lambda_p^{m_p}\{\prod_{j=1}^p(1-\lambda_j)^{\alpha-\frac{n+1}{2}}\}{\rm d}D\tag{16}\end{align*}
where, $\alpha >\frac{n-1}{2}$ and $m_j=\frac{n}{2}-\frac{p+1}{2}+p-k_j$.
\vskip.2cm
In the complex case,
$$\tilde{g}_3(D){\rm d}D=\frac{\tilde{\Gamma}_p(\alpha+p)}{\tilde{\Gamma}_p(n)\tilde{\Gamma}_p(\alpha+p-n)}\frac{\pi^{p(p-1)}}{\tilde{\Gamma}_p(p)}\sum_K(-1)^{\rho_K}\lambda_1^{m_1}...\lambda_p^{m_p}\{\prod_{j=1}^p(1-\lambda_j)^{\alpha-n}\}{\rm d}D\eqno(17)$$
where $\alpha >n-1$ and $m_j=n-p+r_j$ where $r_j$ is defined in (15). Hence, in (16) and (17) we will use the same notation $m_j$ as the exponent of $\lambda_j,j=1,...,p$ with the understanding that in the real case $m_j=\frac{n}{2}-\frac{p+1}{2}+p-k_j$ and in the complex case $m_j=n-p+r_j$. Further, for simplicity, we may write the joint density of the eigenvalues in the real and complex cases as the following:
\begin{align*}
g_3(D){\rm d}D&=c\sum_K(-1)^{\rho_K}\lambda_1^{m_1}...\lambda_p^{m_p}(1-\lambda_1)^{\gamma}...(1-\lambda_p)^{\gamma}{\rm d}D\tag{18}\\
\tilde{g}_3(D){\rm d}D&=\tilde{c}\sum_K(-1)^{\rho_K}\lambda_1^{m_1}...\lambda_p^{m_p}(1-\lambda_1)^{\gamma}...(1-\lambda_p)^{\gamma}{\rm d}D\tag{19}\end{align*}where
$$m_j=\frac{n}{2}-\frac{p+1}{2}+p-k_j,\gamma=\alpha-\frac{n+1}{2},c=\frac{\Gamma_p(\alpha+\frac{p}{2})}{\Gamma_p(\frac{n}{2})\Gamma_p(\alpha+\frac{p}{2}-\frac{n}{2})}
\frac{\pi^{\frac{p^2}{2}}}{\Gamma_p(\frac{p}{2})},j=1,...,p$$
in the real case, and
$$m_j=n-p+r_j,\gamma=\alpha-n,\tilde{c}=\frac{\tilde{\Gamma}_p(\alpha+p)}{\tilde{\Gamma}_p(n)\tilde{\Gamma}_p(\alpha+p-n)}\frac{\pi^{p(p-1)}}{\tilde{\Gamma}_p(p)},j=1,...,p$$
in the complex case, where $r_j$ is defined in (15). Since we have written the joint density of the eigenvalues, both in the real and complex cases, by using the same format, we can use the same procedure to obtain the densities of the largest eigenvalue, smallest eigenvalue etc. Integration over $\lambda_1,...,\lambda_{p-1}$ is needed to obtain the density of the smallest eigenvalue $\lambda_p$. Similarly, integration over $\lambda_p,...,\lambda_2$ is needed to obtain the density of the largest eigenvalue $\lambda_1$. In the complex case, $m_j,j=1,...,p$ are always  positive integers. Hence, integration by parts can get rid off the factor $\lambda_j^{m_j}$. But, if the $m_j$ is large or moderately large then the final expression, even though a finite sum, will be messy. Similarly, when $m_j$ or $\gamma$ in the real or complex case is a positive integer, then integration by parts will eliminate the corresponding factor either $\lambda_j^{m_j}$ or $(1-\lambda_j)^{\gamma}$. But the expressions may become messy when the parameter $m_j$ or $\gamma$ is large or moderately large. Hence, we will consider series expansions which will be valid for the real and complex cases. When the parameters are positive integers, then these series will terminate into a finite sum. Since $0<\lambda_j<1$ the series will converge fast even if it is an infinite series.

\vskip.3cm\noindent{\bf 4.1.\hskip.3cm Exact marginal function of the largest eigenvalue $\lambda_1$ in (18)}

\vskip.3cm
 For both the real and complex cases, whether $\gamma$ is a positive integer or not, let us expand $(1-\lambda_j)^{\gamma}$  to obtain a convenient representation. Note that
 $$(1-\lambda_p)^{\gamma}=\sum_{t_p= 0}^{\infty}\frac{(-\gamma)_{t_p}}{t_p!}\lambda_p^{t_p}.$$
 This will be a finite sum when $\gamma$ is a positive integer. Now, we start integrating from $\lambda_p$ onward.
 $$\int_{\lambda_p=0}^{\lambda_{p-1}}\lambda_p^{m_p}(1-\lambda_p)^{\gamma}{\rm d}\lambda_p=\sum_{t_p=0}^{\infty}\frac{(-\gamma)_{t_p}}{t_p!}\frac{1}{m_p+t_p+1}\lambda_{p-1}^{m_p+t_p+1}.$$ Now, multiply this with $\lambda_{p-1}^{m_{p-1}}(1-\lambda_{p-1})^{\gamma}$ and integrate $\lambda_{p-1}$ from $0$ to $\lambda_{p-2}$, and so on. The final form, denoted by $f_1(\lambda_1)$ is the marginal function corresponding to the largest eigenvalue $\lambda_1$. That is,
 \begin{align*}
 f_1(\lambda_1)&=c\sum_K(-1)^{\rho_K}\sum_{t_p=0}^{\infty}\frac{(-\gamma)_{t_p}}{t_p!}\frac{1}{m_p+t_p+1}\sum_{t_{p-1}=0}^{\infty}\frac{(-\gamma)_{t_{p-1}}}{t_{p-1}!}\frac{1}{m_p+m_{p-1}+t_p+t_{p-1}+2}...\\
 &\times \sum_{t_2=0}^{\infty}\frac{(-\gamma)_{t_2}}{t_2!}\frac{1}{m_p+...+m_2+t_p+...+t_2+(p-1)}\lambda_1^{m_p+...+m_1+t_p+...+t_2+(p-1)}(1-\lambda_1)^{\gamma},\end{align*}
 for $0\le \lambda_1\le 1$ and zero elsewhere. Now, incorporating the remaining factors from (16) and (17), we have the marginal density of $\lambda_1$.

 \vskip.3cm\noindent{\bf 4.2.\hskip.3cm Exact marginal function of $\lambda_p$, the smallest eigenvalue}
\vskip.3cm
In the complex case, $m_j$ is a positive integer for all $j$, and in the real case, $m_j$ is either a positive integer or a half-integer. Since we are integrating out, starting from $\lambda_1,\lambda_2<\lambda_1<1$, we may write, for convenience,
$$\lambda_1^{m_1}=[1-(1-\lambda_1)]^{m_1}=\sum_{t_1=0}^{\infty}\frac{(-m_1)_{t_1}}{t_1!}(1-\lambda_1)^{t_1}.$$
Then,
\begin{align*}
\int_{\lambda_1=\lambda_2}^1\lambda_1^{m_1}(1-\lambda_1)^{\gamma}{\rm d}\lambda_1&=\sum_{t_1=0}^{\infty}\frac{(-m_1)_{t_1}}{t_1!}(1-\lambda_1)^{\gamma+t_1}{\rm d}\lambda_1\\
&=\sum_{t_1=0}^{\infty}\frac{(-m_1)_{t_1}}{t_1!}\frac{1}{\gamma+t_1+1}(1-\lambda_2)^{\gamma+t_1+1}.
\end{align*}
Now, multiply by $\lambda_2^{m_2}(1-\lambda_2)^{\gamma}$ and integrate out $\lambda_2$, and so on. Final result, denoted by $f_p(\lambda_p)$, is the following:

\begin{align*}
f_p(\lambda_p)&=c\sum_K(-1)^{\rho_k}\sum_{t_1=0}^{\infty}\frac{(-m_1)_{t_1}}{t_1!}\frac{1}{\gamma+t_1+1}\sum_{t_2=0}^{\infty}\frac{(-m_2)_{t_2}}{t_2!}\frac{1}{2\gamma+ t_1+t_2+2}...\\
&\times \sum_{t_{p-1}=0}^{\infty}\frac{(-m_{p-1})_{t_{p-1}}}{t_{p-1}!}\frac{1}{(p-1)\gamma+t_1+...+t_{p-1}+(p-1)}\\
&\times \lambda_p^{m_p}(1-\lambda_p)^{p\gamma+t_1+...+t_{p-1}+(p-1)},0\le \lambda_p\le 1
\end{align*}
and this multiplied by the remaining factors from (16) and (17) will be the marginal density of the smallest eigenvalue $\lambda_p$. Observe that if one wishes to compute the marginal density of $\lambda_s$ for any $s$, then integrate out $\lambda_1,...,\lambda_{s-1},\lambda_p,...,\lambda_{s+1}$. Then, multiply by the remaining factors from (16) and (17). By integrating out $\lambda_1,...,\lambda_{s-}$, one obtains the joint marginal function of $\lambda_s,...,\lambda_p$, and so on.

\vskip.3cm\noindent{\bf Acknowledgement:} The authors wish to thank the referee for the extensive, thoughtful and very valuable comments and suggestions which enabled the authors to improve the presentation of the paper.
\vskip.3cm\noindent{\bf Author contributions:} All authors contributed equally go this study. All authors have read and agreed to the final version of the manuscript.
\vskip.3cm\noindent{\bf Funding:} This research received no external funding.
\vskip.3cm\noindent{\bf Data Availability Statement:} No new data were created or analyzed in this study. Data sharing is not applicable to this article.
\vskip.3cm\noindent{\bf Conflicts of Interest:} The authors declare no conflict of interest.

\begin{center}{\bf References}\end{center}

\vskip.3cm\noindent [1]~~F.F. Deppisch (2019): {\it A Modern Introduction to Neutrino Physics}, Morgan $\&$ Claypool Publishers, San Rafael, CA, USA.

\vskip.3cm\noindent [2]~~L. Oberauer, A. Ianni, and A. Serenelli (2020): {\it Solar Neutrino Physics: The Interplay between Particle Physics and Astronomy}, Wiley-VCH, Weinheim, Germany.
\vskip.3cm\noindent [3]~~N. Haba and H. Murayama (2001): Anarchy and Hierarchy, {\it Physical Review} {\bf D 63}, 053010, arXiv:hep-ph/0009174.
\vskip.3cm\noindent [4]~~T. Yanagida (1979): Horizontal Symmetry and Mass of the Top Quark, {\it Physical Review} {\bf D 20}, 2986.
\vskip.3cm\noindent [5]~~J.-F. Fortin, N. Glasson, and L. Marleau (2016): Probability density function for neutrino masses and mixings, {\it Physical Review}, {\bf D 94},(115004-1)-(11500-13).
\vskip.3cm\noindent [6]~~N. Scafetta (2010): {\it Fractal and Diffusion Entropy Analysis of Time Series: Theory, concepts, applications and computer codes for studying fractal noises and L\'evy walk signals}, VDM Verlag Dr. Mueller, Saarbruecken.
\vskip.3cm\noindent [7]~~A.M. Mathai and H.J. Haubold (2018): {\it Erd\'elyi-Kober Fractional Calculus from a Statistical Perspective, Inspired by Solar Neutrino Physics}, Springer Briefs in Mathematical Physics 31, Springer Nature, Singapore.
\vskip.3cm\noindent [8]~~H.J. Haubold and A.M. Mathai (2024): Does SuperKamiokande Observe L\'evy Flights of Solar Neutrinos?, https://doi.org/10.48550/arXiv.2405.11057.
\vskip.3cm\noindent [9]~~N. Haba, Y. Shimizu, and T. Yamada: Neutrino mass square ratio and neutrinoless double-beta decay in random neutrino mass matrices, {\it Progress in Theoretical Physics}, 2023, 023B07.
\vskip.3cm\noindent [10]~~A.M. Mathai (1997): {\it Jacobians of Matrix Transformations and Functions of Matrix Arguments}, World Scientific, New York.
\vskip.3cm\noindent [11]~~A.M. Mathai, S.B. Provost, and H.J. Haubold (2022): {\it Multivariate Statistical Analysis in the Real and Complex Domains}, Springer Nature, Switzerland.
\vskip.3cm\noindent [12]~~J.-F. Fortin, N. Giasson, L. Marleau, and J. Pelletier-Dumont (2020): Mellin transform approach to rephasing invariants, {\it Physical Review}, {\bf D 102}, 036001.
\vskip.3cm\noindent [13]~~A.M. Mathai and SD.B. Provost (2024): The exact densities of the eigenvalues of Wishart and matrix-variate gamma and beta random variables, {\it Mathematics},2024,12,2427 doi.org/10.3390/math12132427.

\end{document}